\documentstyle[multicol,epsfig,aps,prl]{revtex}

\ifpreprintsty
 \def\multb{ }
 \def\multe{ }
 \else
 \def\multb{ \begin{multicols}{2}}
 \def\multe{ \end{multicols}}
 \fi
\begin{document}
\draft
\title{How to calculate the degree of spin polarization in ferromagnets.}
\author{I.I. Mazin}
\address{Code 6391, Naval Research Laboratory, Washington, DC 20375}
\date{December 17, 1998}
\maketitle

\begin{abstract}
Different ways to define and  calculate the degree of spin
polarization in a ferromagnet are discussed, particularly with respect to
spin-polarized tunneling and Andreev reflection at the boundary between
superconductor and ferromagnet. As an example, the degree of spin
polarization for different experiments in Fe and Ni is calculated in the
framework of the local spin density approximation (LSDA) and used to
illustrate the difference between various definition of spin
polarization.
\end{abstract}

\multb
Although solid state physicists use the notion of a degree of 
 spin polarization (DSP) of a ferromagnet (FM) rather often, it is
not well defined. While the total magnetic moment is uniquely defined as the
difference between the number of spin up and spin down electrons, it
 tells us little about how much do electrons with the different spins
contribute to transport properties. In view of the
growing number of experiments probing various aspects of spin polarization%
\cite{meservey94}, it becomes increasingly more important to be able to
calculate the DSP in the framework of the conventional band theory (and
eventually beyond it). Importantly, the DSP can be defined in several
different ways. Therefore it is crucial, in order to compare the
calculations with the experimental data, to make sure that a proper
definition of the DSP is used. In particular, spin-polarized tunneling in
various forms\cite{meservey94}, including Andreev reflection\cite{sci}, 
provides valuable information about the spin dependence of the
electronic structure, but this information may be obscure and not very
useful unless the measurements are backed by the calculation {\it %
appropriate for the experiment in question}.

Let us consider an extreme example, the so called half-metallic magnets,
for instance CrO$ _{2}$. Such
systems do not have any electrons at the Fermi level in one of the two spin
channels; they are metals in one spin channel and
semiconductors in the other. 
Half metals have 100\% spin polarization according to any sensible
definition. On the other hand, for a regular magnetic metal, which has Fermi
surfaces in both spin channels, it is not obvious {\it a priori }how to
define the degree of spin polarization.

The most natural definition, and probably the one most often used is 
$
P=(N_{\uparrow }-N_{\downarrow })/(N_{\uparrow }+N_{\downarrow }),
$
where $N_{\uparrow (\downarrow )}$ is the density of electronic states (DOS)
at the Fermi level, defined as ($\hbar \equiv 1$)
\begin{equation}
N_{i} =\frac{1 }{(2\pi
)^{3}}\sum_{\alpha }\int \delta (E_{{\bf k}\alpha i})d ^3 k 
=\frac{1 }{(2\pi )^{3}}\sum_{\alpha }\int \frac{dS_{F}}{v_{{\bf k}%
\alpha i}},  \label{N}
\end{equation}
and $E(v)_{{\bf k}\alpha i}$ is the energy (velocity) of an electron in the
band $\alpha $ with spin $i(=\uparrow or\downarrow )$ and the wave vector $%
{\bf k.}$ A typical experiment
that can probe $P_{N}$ is spin-polarized photoemission. This definition of
the DSP may be called ``$N$''-definition, $P_{N}$, and its usefulness is
limited by the fact that the transport phenomena usually are {\it not}
defined by the DOS alone.
This is particularly true for materials which have both
heavy $d$-electrons and light $s$-electrons at the Fermi level (a good
example is Ni). While the DOS is mostly defined by the former, the electric
transport is primarily due to the fast $s$-electrons (cf. a semiempirical
recipe of defining DSP via partial $s$-DOS in transition metals, \cite{sd}). 

Classical Bloch-Boltzmann theory of electric transport in metals\cite{allen}
 lets us
separate the current of the spin-up electrons and the current of the
spin-down electrons, and to define
DSP {\rm via} the current densities $J_{\uparrow (\downarrow )}$ as $%
(J_{\uparrow }-J_{\downarrow })/(J_{\uparrow }+J_{\downarrow }),$
$J_{\uparrow (\downarrow )}\propto \left\langle Nv^{2}\right\rangle
_{\uparrow (\downarrow )} \tau _{\uparrow (\downarrow )}$. Assuming
the same relaxation time $\tau $ for both spins, this definition leads to
the ``$Nv^{2}$'' DSP: 
\begin{equation}
P_{Nv^{2}}=(\left\langle Nv^{2}\right\rangle _{\uparrow }-\left\langle
Nv^{2}\right\rangle _{\downarrow })/(\left\langle Nv^{2}\right\rangle
_{\uparrow }+\left\langle Nv^{2}\right\rangle _{\downarrow }),
\end{equation}
where $\left\langle Nv^{2}\right\rangle _{\uparrow (\downarrow )}$ is
defined as 
\begin{eqnarray}
\left\langle Nv^{2}\right\rangle _{i} &=&
{(2\pi )^{-3}}\sum_{\alpha
}\int v_{{\bf k}\alpha i}^{2}\delta (E_{{\bf k}\alpha i})d ^3 k \nonumber
\\
&=&{(2\pi )^{-3}}\sum_{\alpha }\int {v}_{{\bf k}\alpha
i} {dS}_{F}.  \label{Nv2}
\end{eqnarray}
This quantity is sometimes denoted as $\left( \frac{n}{m}\right) _{{\rm eff}}
$ and is proportional to the contribution of the corresponding electrons to
the plasma frequency (see, e.g. Ref.\cite{allen}). If spin-dependent or
spin-flip scattering is present, the total current in each spin channel
depends on the characteristics of both spin subsystems, and the expression
for the DSP becomes very complicated.

Unfortunately, it is hardly possible to measure $%
J_{\uparrow }$ and $J_{\downarrow }$ separately. A typical experiment
involves spin-polarized tunneling between a FM and another material. In
particular, it is possible to measure tunneling currents separately for both
spin polarizations for a ferromagnet/superconductor contact.
The question arises, whether the DSP measured in such a way is $P_{N}
$ or $P_{Nv^{2}}?$
To answer this, let us start from the simplest case,
a ballistic contact with no barrier, and neglecting possible mismatch of
the  Fermi velocities at the contact. We 
 repeat the original Sharvin\cite{sharvin65}
semiclassical derivation, but allow for
arbitrary Fermiology. Following Sharvin, we assume that an electron 
going through the contact experiences the acceleration by the electric field
so
that its energy increases by $eU.$ Assuming
that the field changes the electron's quasimomentum from $\hbar {\bf k}$ to $%
\hbar {\bf k}^{\prime },$ we find that the phase space for this process is
defined at $T=0$  by the factor 
\begin{equation}
\theta (E_{{\bf k}^{\prime }})\theta (-E_{{\bf k}})=\theta (E_{{\bf k}%
}+eU)\theta (-E_{{\bf k}})=eU\delta (E_{{\bf k}}).
\end{equation}
The fraction of electrons with a given {\bf k} that can reach the contact in
a unit time is $v_{x}A$
(the contact plane is perpendicular to $x$ and
$A$ is the area of the contact). The total
current is 
\begin{eqnarray}
I&=&\frac{e^{2}UA}{(2\pi )^{3}}\sum_{\alpha }\int_{v_{x}>0}v_{x}\delta (E_{%
{\bf k}\alpha })d{\bf k=}{e^{2}UA}\left\langle
Nv_{x}\right\rangle=\nonumber\\
&=&\frac{e^{2}UA}{(2\pi )^{3}}\sum_{\alpha }\int_{v_{x}>0}v_{x}
\frac{dS_F}{v}=e^{2}UAS_x,  \label{sharv}
\end{eqnarray} where $S_x$ is the area of the projection of the
Fermi surface onto the interface plane.
For a Fermi sphere this reduces to the Sharvin result. Correspondingly,
we arrive at the third, ballistic definition of spin polarization,
$P_{Nv}$.

The next simplest model is that of a specular ($%
\delta $-function) tunneling barrier with an $f/s$ Fermi
velocity mismatch. Here we need to take into account,
in addition to the $(v_x/v)$ factor, a finite 
barrier transparency. It depends (nonpolinomially) on the Fermi velocities
\cite{mazin95}: 
\begin{equation}
D=\frac{v_{xf}v_{xs}}{(v_{xf}+v_{xs})^{2}+W_{i}^{2}},
\end{equation}
where
$v_{s(f)}$ is the Fermi
velocity in the superconductor(ferromagnet),
and $W$ is the strength of the barrier, $%
V(x)=W\delta (x)$ (for a one-band isotropic material $W$ is related with the
dimensionless parameter $Z$ of Ref.\cite{blonder82} as $Z=W/v_{F}).$
Tunneling current is thus proportional to
\begin{equation}
\int_{v_{xf}>0}Dv_{xf}\frac{dS_{F}}{v_{f}}\propto \int_{v_{xf}>0}%
\frac{v_{xf}^{2}}{(v_{xf}+v_{xs})^{2}+W^{2}}\frac{dS_{F}}{v_{f}}.\label{tun}
\end{equation}
In the large $W$ limit this reduces to $\left\langle Nv^{2}\right\rangle $
and the measured DSP is $P_{Nv^{2}}.$ However, in the high
transparency limit one cannot give a simple answer.

It was recently suggested\cite{de95} that Andreev reflection at the
interface between a FM and a superconductor (SC) can be used for direct probing
of DSP. The idea is simple: the Andreev reflection can be visualized as two
currents of electrons with the opposite spins flowing inside the normal
metal towards its interface with a SC. At the interface (more
precisely, within the coherence length from the interface) the two currents
recombine creating the current of Cooper pairs. One can also say that to
form a Cooper pair an electron has to leave behind a hole, creating an
additional hole current flowing in the opposite direction. In a paramagnetic
(or antiferromagnetic with time reversal symmetry) metal both currents
are the same, so one observes in the superconducting state the 100\%
increase in the net current over the normal state. De Jong and Beenacker\cite
{de95} suggested that in a FM the total Andreev current is defined by that
spin channel where the normal-state current is smaller, because the excess
electrons in the other channel will not find partners to form Cooper pairs
with. This statement was quantified in Ref.\cite{de95} {\it via} the number
of spin-up and spin-down conductance channels, which they denoted as $%
N_{\uparrow (\downarrow )},$ thus arriving at an expression for the ratio of
the current in the superconducting and the normal state as
\begin{equation}
I_s/I_n=4\min
(N_{\uparrow },N_{\downarrow })/(N_{\uparrow }+N_{\downarrow }).
\label{BJ}
\end{equation}
 The number
of conductance channels cannot be directly evaluated. Besides, this
expression can mislead the reader into a belief that the DSP measured
through Andreev reflection is  $P_{N}$ (i.e., defined by the DOS).

In fact, Andreev reflection at a FM/SC contact has been
recently attracting substantial theoretical interest 
\cite{misc}. This interest so far concentrated on such aspects as
the symmetry of the order parameter, Fermi velocity mismatch,
and generalization of the BTK formula\cite{blonder82} onto spin-polarized
case\cite{sci,misc}.
In terms of electronic structure, however, all the work was limited
to the parabolic bands/spherical Fermi surface model. While revealing
important
fundamental physics, such approach is of limited practical importance,
because in real ferromagnetic material this approximation is absolutely
inacceptable. In this Letter we, on the other hand, focus on the
band structure effects in spin polarization, and in this context we
need a better definition for the ``number of conductance channels''.

Comparing Eq.\ref{BJ} with Eqs.\ref{sharv} and \ref{tun}, we observe 
that the DSP for Andreev reflection should be defined as
either $P_{Nv^{2}}$, for a large
barrier and/or diffusive current, or $P_{Nv}$, for low resistance
ballistic contacts. This is, however, only the first approximation,
while full expressions should include Fermi surface averages of 
more complicated functions of $v_F$.
For one particular case, a fully ballistic
(Sharvin) Andreev reflection, an explicit formula,
reflecting the physics suggested by De Jong and Beenacker,
can be derived, which is both suitable for band structure
calculations and also quite illustrative.

\begin{figure}[tbp]
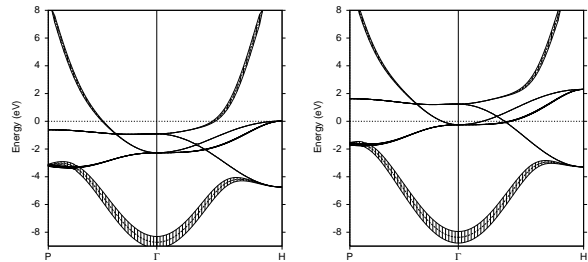

\centerline{\epsfig{file=febndup.epsi,height=0.45\linewidth,angle=-90}
\epsfig{file=febnddn.epsi,height=0.45\linewidth,angle=-90}}
\vspace{0.1in} \setlength{\columnwidth}{3.2in} \nopagebreak
\caption{Fe bands in the two spin channels. The linewidth is proportional to
the partial $s$-character in each state. }
\label{Febands}
\end{figure}
\begin{figure}[tbp]
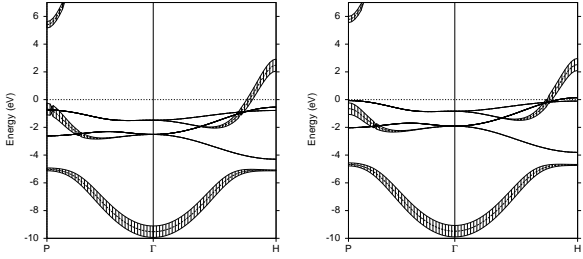

\centerline{\epsfig{file=nibndup.epsi,height=0.45\linewidth,angle=-90}
\epsfig{file=nibnddn.epsi,height=0.45\linewidth,angle=-90}}
\vspace{0.1in} \setlength{\columnwidth}{3.2in} \nopagebreak
\caption{The same as Fig.\ref{Febands}, for Ni. }
\label{Nibands}
\end{figure}

De Jong and Beenaker treated incoming electrons
and reflected holes as two separate currents. In purely ballistic
regime, however, one has to take into account energy and momentum
conservation (parallel to interface) for each individual reflected
hole, so that $P_{Nv}$, with its independent averaging over each 
spin channel, does not necessarily correctly describe
observable polarization.
This statement may be quantified as follows:
Consider  an incoming electron  with
a momentum ${\bf k=(k}_{\parallel },k_{x})$, which is reflected as
 a hole with the momentum 
${\bf q=(q}_{\parallel },q_{x})$ in the other spin subband; ${\bf q}%
_{\parallel }={\bf k}_{\parallel },$ $E_{{\bf k}\uparrow }=E_{{\bf q}%
\downarrow }.$ This fixes for each ${\bf k}$ a countable number (one, two,
etc) of wavevectors ${\bf q}$ satisfying this condition. For simplicity we
will assume for the moment that there is only one such ${\bf q}$ and will
denote it ${\bf \tilde{q}.}$ This adds an additional constraint to Eq.\ref
{sharv}, so that instead we have, omitting summation over the band indices 
$\alpha $,
\[
I=\frac{e^{2}UA}{(2\pi )^{6}}
\int_{v_{x}>0}\delta ({\bf q}_{\parallel }-{\bf %
k}_{\parallel })\delta (q_{x}-\tilde{q}_{x})v_{x}\delta (E_{{\bf k}})d ^3k
d ^3q.\nonumber
\]
This can be rewritten in a symmetric way like
\multe
\begin{equation}
I=(2\pi )^{-6}{e^{2}UA}
\int_{v_{x},u_{x}>0}d^{3}{ k}d^{3}{ q}%
\int d^{2}{\bf a[}\delta ({\bf q}_{\parallel }-{\bf a})\delta ({\bf k}%
_{\parallel }-{\bf a})][u_{x}\delta (E_{{\bf q}})][v_{x}\delta (E_{{\bf k}})]%
{\bf ,}  \label{a}
\end{equation}
\multb
\vspace{-5mm}
where we introduced a 2D vector ${\bf a}$, transformed the $\delta$-function
of $q$ into a $\delta$-function of $E_{\bf q}$, and introduced
${\bf u=}\partial E_{%
{\bf q}}/\partial {\bf q.}$ We can further rewrite Eq.\ref{a} as 
\begin{equation}
I=e^{2}UA\int d^{2}{\bf a}F_{\uparrow }({\bf a})F_{\downarrow }({\bf a)=}%
(2\pi )^{-3}{e^{2}UAS},
\end{equation}
where $F({\bf a})=(2\pi )^{-3}\int_{v_{x}>0}\delta ({\bf k}%
_{\parallel }-{\bf a})v_{x}\delta (E_{{\bf k}})$ is half (because $v_{x}>0)$
the number of crossings of the line ${\bf k_{\parallel }=a}$ with the Fermi
surface for a given spin. Using again projections of 
the Fermi surface for either spin
onto the interface plane, one can define $S$ as the
overlap area of the spin-up and spin-down projections. This current can be
expressed in terms of a spin polarization as in Eq.\ref{BJ}, thus giving
yet another, ``ballistic Andreev'' definitinion of spin 
polarization, similar, but not the same as the ``$Nv$'' definition,
$P_{\rm b.A.}\neq P_{Nv}$ (they are equal only if the Fermi surface
projection for one spin is entirely contained in that for the other spin). 

So we observe that an experiment would probe different DSP's depending on
the length scale of the problem, which is defined by the size of the contact
and the length at which the voltage drops, and how it compares with the mean
free path. Furthermore, the transparency of the barrier can also influence
the measured DSP. In the pure ballistic limit the DSP is related to the
average Fermi velocity,
while in the purely diffusive regime it is defined by the average
squared Fermi velocity\cite{rev}.
One may ask, why DOS is so often used as a measure
of spin polarization, even though such definition is irrelevant for transport
properties? The answer is that the most common way to perform tunneling or
similar experiments is to follow the details of the contact conductance as a
function of applied voltage. Probably the most spectacular and fruitful
application of this technique is the tunneling spectroscopy of
superconductors. In such case the characteristic scale for the voltage
change is the superconducting gap and Debye temperature. The normal state
electronic structure does not change over such a small energy range, so the
only important factor is the variation of the superconducting DOS with
energy. The normal state DOS and velocity can be assumed constant and
factored out. Of course, it is not the case when two different sheets of the
Fermi surface, as in Ref.\cite{mazin95}, or two different spin channels, are
compared.

\begin{figure}[tbp]
\centerline{\epsfig{file=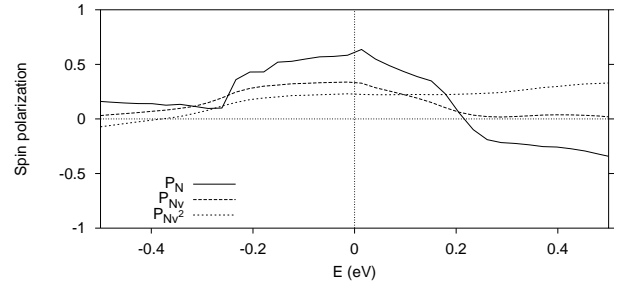,width=0.95\linewidth}}
\vspace{0.in} \setlength{\columnwidth}{3.2in} \nopagebreak
\caption{Degree of spin polarization for Fe, 
calculated as $P_N$, $P_{Nv}$, and $P_{Nv^2}$}
\label{FeDSP}
\end{figure}

\begin{figure}[tbp]
\centerline{\epsfig{file=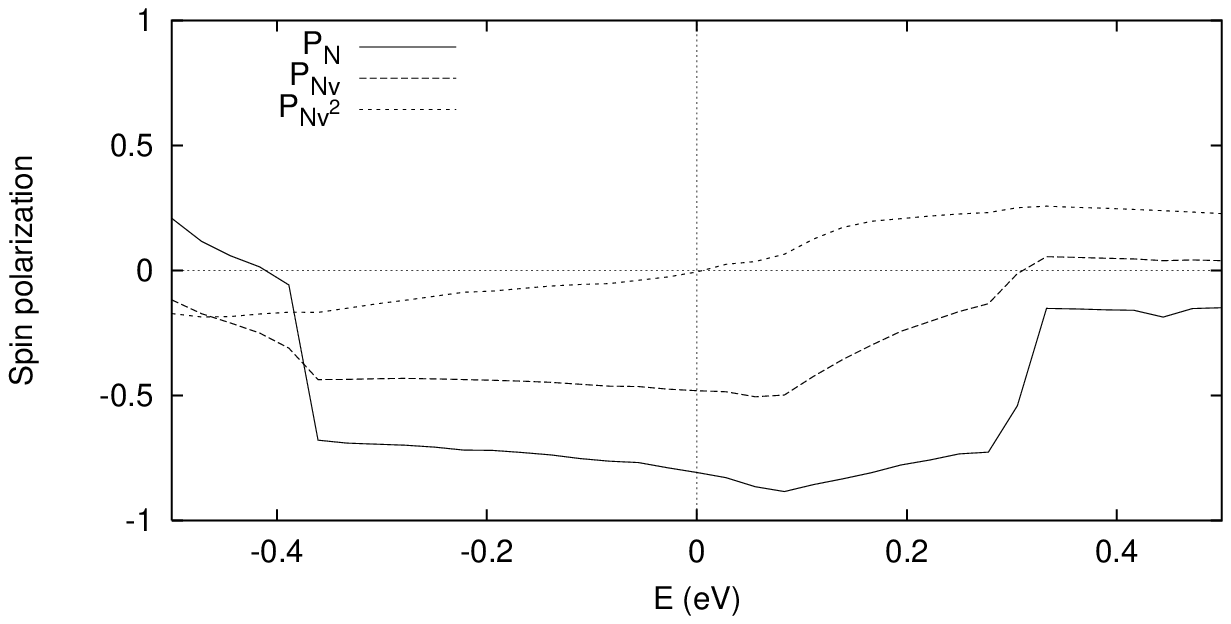,width=0.95\linewidth}}
\vspace{0.in} \setlength{\columnwidth}{3.2in} \nopagebreak
\caption{The same as Fig.\protect\ref{FeDSP}, for Ni.  }
\label{NiDSP}
\end{figure}

Importantly, $P_{Nv^{2}},$ $P_{Nv}$  and $P_{N}$ are entirely different in
real materials (Figs. \ref{FeDSP},\ref{NiDSP}). The reason is (and the ``$s$%
-DOS recipe'' works for the same reason) that in transition metals one can
often distinguish the pieces of the Fermi surface that are predominantly $d$
in character and the pieces that are mostly $s$. The former have low Fermi
velocity and, according to Eq.\ref{N}, are responsible for the most of the
DOS. The latter have high Fermi velocity and provide the main contribution
to Eq.\ref{Nv2}. The larger the anisotropy of the Fermi velocity (angular
anisotropy, in principle, works in the same way as interband one), the
larger is the difference between $P_{Nv^{2}}$, $P_{Nv}$ and $P_{N}$.

\begin{figure}[tbp]
\centerline{\epsfig{file=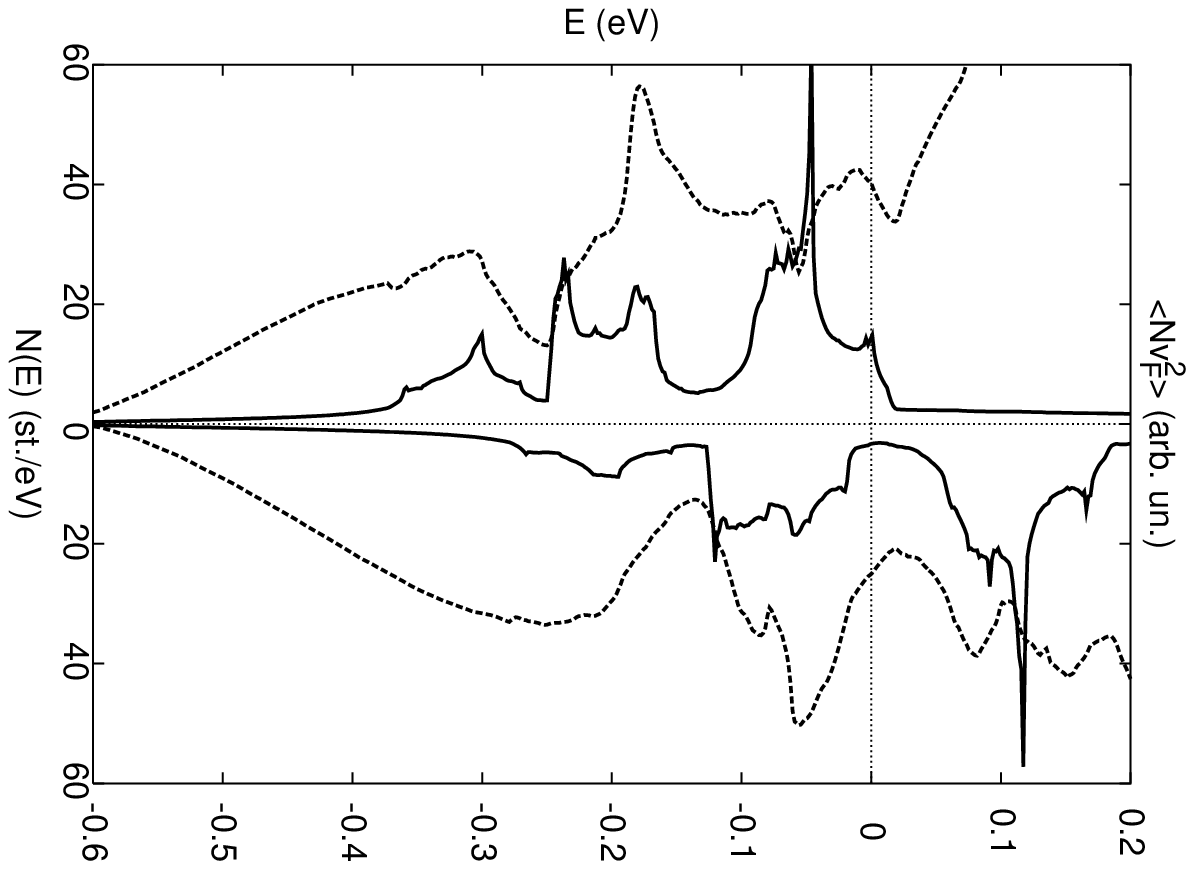,height=0.45\linewidth,angle=90}
\epsfig{file=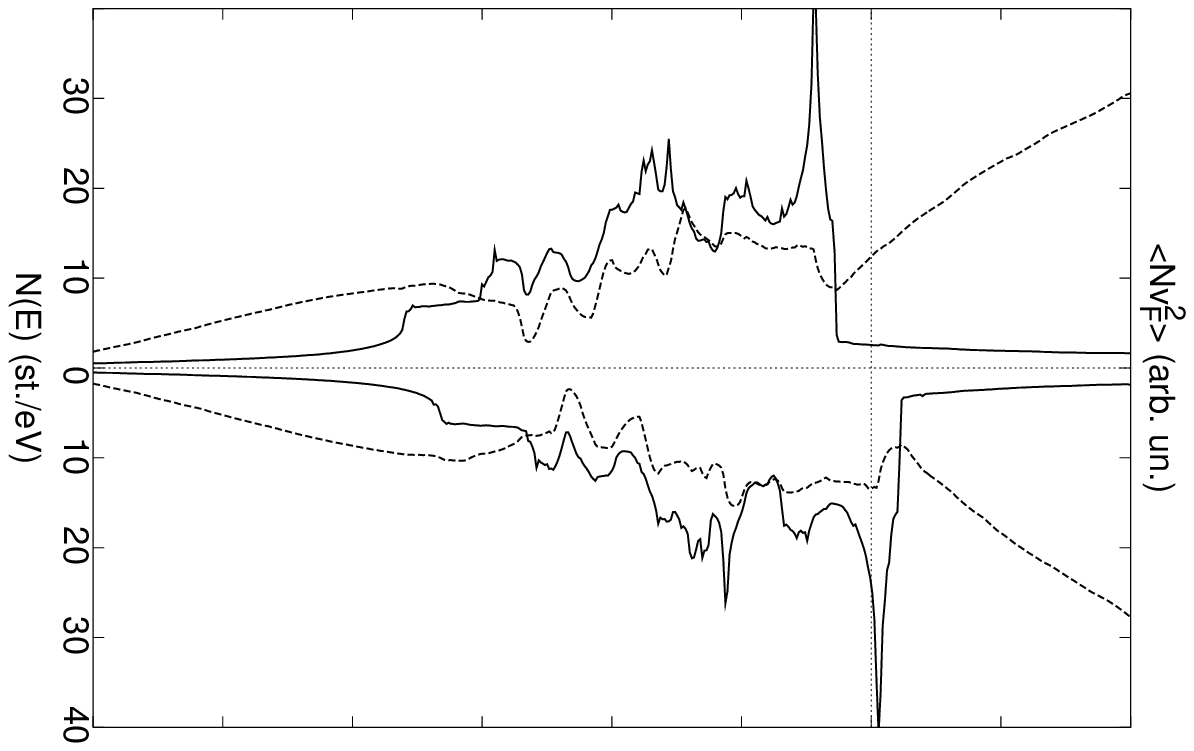,height=0.45\linewidth,angle=90}}
\vspace{0.in} \setlength{\columnwidth}{3.2in} \nopagebreak
\caption{Density of states, $N(E)$ (solid line) and 
$\left( \frac{n}{m}\right) _{{\rm eff}
}(E)= \langle Nv_F^2\rangle (E) $ (dashed line), for Fe (left panel)
and Ni (right panel).}
\label{FeDOS}
\end{figure}

To illustrate this difference in real materials we present here Linear
Muffin-Tin Orbitals\cite{OKA75} LSDA calculations of the corresponding
quantities in Fe and Ni. It is instructive to start from the band structure
itself (Figs.\ref{Febands},\ref{Nibands},\ref{FeDOS}). The ``fat'' bands in
these figures correspond to the states with substantial $sp$ character. One
immediately notices a qualitative difference between Fe and Ni: In the
former all bands in both spin channels are heavily hybridized at the Fermi
level, and one cannot convincingly classify bands as ``predominantly $sp$''
and ``predominantly $d$''. In the latter the $d$ band is so deep that one
can single out an $sp$-like pocket in the spin-up channel (in Fig. \ref
{Nibands}, the band crossing Fermi level between $\Gamma $-H), and $d$-like
pockets (in Fig. \ref{Nibands}, near H). The Fermi surface in the spin-down
channel is entirely $sp$-like. This is similar to paramagnetic Pd\cite
{mazin84}, which is the only $4d$ metal where transport properties can be
described by the $s-d$ scattering model.

In general a large  $P_{N}$  may be due to any of the two reasons: either
the {\it areas} of the up- and down-Fermi surfaces are different, while
velocities may be similar, or the areas are not too different, but the Fermi 
{\it velocity }for one spin channel is much smaller than for the other. In
the former case the additional factors of $v_{F}$ or $v_{F}^{2}$ may change
the DSP somewhat, but qualitative changes, or, in an extreme case,
the sign change,
are unlikely. If, however, ``light'' and ``heavy'' electrons are present,  $%
P_{N}$ is dominated by the ``heavy'' pockets, and $P_{Nv^{2}}$ by the
``light'' ones, so the two DSP's are likely to be very different and
possibly have opposite signs.

The first situation is realized in Fe. One indeed can see that $N(E)$ and $%
\langle Nv_{F}^{2}\rangle (E)$ behave  similarly (Fig.\ref{FeDOS}).
Correspondingly, the difference between different DSP's is only moderate
(Fig. \ref{FeDSP}). On the other hand, in Ni the DSP essentially drops to 
zero when the factor $v_{F}^{2}$ is included. Interestingly, most
experimental results\cite{meservey94} indicate that the DSP observed in
tunneling is {\it positive} and not too small ($>20$\%), in other words, the
effect described above appears to be even stronger
 in reality than in band
structure calculations. This is also to be expected: the local density
approximation (LDA) has a tendency to underlocalize $d$ electrons in
transition metals. For instance, in Cu the fully occupied $d$-band appears
in the calculations about 0.4 eV higher than in experiments, and is also too
wide\cite{IEG}. Similarly, LDA underlocalization of the $d$ electrons in Ni
leads to an overestimation of the $d$ bandwidth and of the exchange
splitting (by approximately a factor of two). As a result, the separation of
carriers into $sp$-like and $d$-like in Ni should be even more pronounced
compared to
 LDA calculations, and thus the effect of Fermi velocity on DSP even
stronger. This leads, in turn, to the DSP sign reversal, observed in
tunneling experiments.

The author is thankful to J. Byers, E. Demler, A. Golubov, B. Nadgorny,
M. Osofsky, and R. Soulen for many useful 
discussions on experimental and theoretical aspects of the spin-dependent
Andreev reflection.
This work was supported by the ONR. 

\begin{references}
\bibitem{meservey94}  R. Meservey and P.~M. Tedrow, Phys.\ Rep.\ {\bf 238},
173 (1994).

\bibitem{sci}  R.J. Soulen {\it et al.}, Science, {\bf 282}, 5386, (1998).

\bibitem{sd}  The ``$s$-recipe'' goes back to late sixties (see, $e.g.,$
J.B. Gadzuk, Phys. Rev. {\bf 182}, 416, 1969), when it was first pointed out
that the tunneling matrix element for $s$-electrons is much larger than that
for $d$-electron. Although later the correct factor of $v_{F}^{2}$ was
identified in many papers (see, $e.g.,$ Z. Yusof {\it et al}. Phys. Rev. 
{\bf B58}, 514, 1998), it is still common to discuss tunneling in
qualitative ``$s$ {\it vs d}'' terms.

\bibitem{allen}  P.B. Allen, Phys. Rev.  {\bf B17}, 3725 (1978).

\bibitem{sharvin65}  Yu. V. Sharvin, ZhETP {\bf 48}, 984 (1965); Sov. Phys.
- JETP {\bf 21}, 655, (1965).

\bibitem{mazin95}  I.~I. Mazin, A.~A. Golubov, and A. Zaikin, Phys.\ Rev.\
Lett.\ {\bf 75}, 2574 (1995).

\bibitem{blonder82}  G.~E. Blonder, M. Tinkham, and T.~M. Klapwijk, Phys.\
Rev.\ B {\bf 25}, 4515 (1982).

\bibitem{de95}  M.~J.~M. de~Jong and C.~W.~J. Beenakker, Phys.\ Rev.\ Lett.\ 
{\bf 74}, 1657 (1995).

\bibitem{misc} 
J.-X. Zhu, B. Friedman, and C. S. Ting, Phys. Rev. {\bf B59}, 9558 (1975);
I. Zutic and O. T. Valls, cond-mat/9808285;
R. L. Merrill and Q. Si, cond-mat/9901004, and others.

\bibitem{rev} See A. G. M. Jansen {\it et al}, J. Phys. C. {\bf 13}, 6073 (1980)
for a general
discussion of the the diffusive and ballistic regime in point
contact spectroscopy.

\bibitem{OKA75}  O.K. Andersen, Phys. Rev. {\bf B12}, 3060 (1975); O.K.
Andersen and O. Jepsen, Phys. Rev. Lett. {\bf 53}, 2571 (1984).

\bibitem{mazin84}  I.~I. Mazin, E.~M. Savitskii, and Y.~A. Uspenskii, J. Phys. {\bf F14%
}, 167 (1984).

\bibitem{IEG}  Theory of the inhomogeneous electron gas, Ed. by S. Lundqvist
and N.H. March, Plenum, 1983.
\end{references}


\multe
\end{document}